\documentclass[12pt]{article}
\usepackage{graphics}
\usepackage{amssymb} 

\author{{\sc K.-H. Rehren} \\
Institut f\"ur Theoretische Physik, Universit\"at G\"ottingen
(Germany)}
\title{\vskip-10mm \bf Chiral Observables and Modular Invariants}
\def\sig{{\sigma}} \def\eps{\varepsilon} \def\a{\alpha} \def\b{\beta} 
\def\g{\gamma} \def\iota{\imath} 
\def\sig{\sigma} \def\lam{\lambda}
\def\inv{{^{-1}}}  \def\comp{{\scriptstyle\circ}} 
 \def\max{^{\rm max}} \def\ext{^{\rm ext}}
\def\stat{^{\rm stat}}
\def\Supp{\hbox{\rm Supp}} \def\Spec{\hbox{\rm Spec}}
\def\frac#1#2{{#1\over#2}}  
\def\eins{\hbox{\rm 1}\mskip-4.4mu\hbox{\rm l}}
 \def\id{{\rm id}} 
\def\ZZ{\mathbb{Z}}  \def\RR{\mathbb{R}} 
 \def\MM{\mathbb{M}} \def\CM{\widetilde \MM} 
\def\qed{\hspace*{\fill}$\square$} \def\SL{\hbox{SL}} \def\PSL{\hbox{PSL}}

\begin{document}
\renewcommand{\today}{}

\maketitle 
{\bf Abstract:} {\small Various definitions of chiral observables in a
  given M\"obius covariant two-dimensional (2D) theory are shown to be
  equivalent. Their representation theory in the vacuum Hilbert space
  of the 2D theory is studied. It shares the general characteristics
  of modular invariant partition functions, although $\SL(2,\ZZ)$
  transformation properties are not assumed. First steps towards
  classification are made. }  

\vskip8mm
\section{Introduction}
The program of classification of modular invariant partition functions
in 2D conformal quantum field theory (see below for more details) has 
seen steady progress since the original A-D-E classification for SU(2)
theories \cite{CIZ}. Apart from explicit classifications for new models
\cite{G}, classification theorems have been established for the 
general case \cite{MS,BE}. Yet, the feeling persists that the full 
depth of the problem has not yet been sounded. 

It is the intention of the present note to show that general
classification theorems of a very similar nature can be derived in a 
setting which does not refer to modular transformations of Gibbs 
states at all. Our statements are on the decomposition (described by 
a ``coupling matrix'') of the vacuum representation of a conformal 2D 
quantum field theory upon restriction to its chiral observables. They 
can be considered with a different perspective as statements on the 
possible 2D extensions of given left and right chiral algebras.
Our mathematical tool is the structure theory of subfactors applied to
the inclusion of local algebras of chiral observables into local algebras 
of 2D observables.

Note that a modular invariant partition function is also described by
a coupling matrix which is usually also interpreted as a chiral 
decomposition of a 2D vacuum representation. But the classification method 
based on arithmetic properties of the representation matrices $S$ and
$T$ of the $\SL(2,\ZZ)$ generators is entirely different and does not 
rely on this interpretation. In fact, there seem to be exotic (accidental?) 
modular invariants which do not derive from a 2D theory \cite[III]{BE}.

In contrast to the modular invariants program, we make only rather 
general structural assumptions on the theory under consideration. We
put the emphasis on the local structure \cite{VK}, rather than the 
accidental Lie algebra structure of chiral observables. Thus we avoid
the, somewhat artificial, restriction to chiral algebras which are 
related to affine Lie algebras because these are the only ones for 
which Gibbs functionals ${\rm Tr}_\pi\,e^{-\b L_0}$ (``characters'') 
are known \cite{KP}. 
Likewise, the problem that for most $W$-algebras it is not clear on 
which suitable set of ``zero mode quantum numbers'' for chiral Gibbs 
functionals the modular group should act, does not pose itself in 
our approach. 

Furthermore, we do not {\em assume} that the left and right chiral 
observables are isomorphic, nor that they have isomorphic fusion of
their superselection sectors. Instead, we shall {\em derive} that the
maximal (see below) chiral observables automatically possess 
sectors with identical fusion rules.

To be sure, it is not our intention to depreciate the modular point of
view at all. On the contrary, the $\SL(2,\ZZ)$ symmetry between high and low
temperature Gibbs states is one of the most fascinating features of
chiral models which calls for a sound physical understanding. Indeed, 
there are general arguments, with reasonable assumptions, in favour of
a modular transformation law which generalizes the one for affine Lie 
algebras \cite{KP} as conjectured in \cite{V}. E.g., Cardy \cite{C} 
argues with transfer matrix methods and invariance under global 
resummations in lattice models before the continuum limit is taken, 
and Nahm \cite{N} exploits the operator product algebra of the Schwinger 
functions to show that Gibbs states transform into Gibbs states. 
None of these, however, provides a completely satisfactory 
explanation in terms of the real time local quantum field theory.

On the other hand, the modular group $\SL(2,\ZZ)$ plays a fundamental
role even without any Gibbs functionals to act on by a modular transformation 
of the temperature. Namely, the general theory of superselection sectors 
collects monodromy data of braid group statistics in numerical 
matrices $S\stat$ and $T\stat$, and as a ``maximality'' feature 
of braid group statistics, these matrices represent 
the modular group \cite{Pal,FG,MM}. In models where both concepts are
defined, one has $S=S\stat$ and $T=T\stat$.
E.g., the Kac-Peterson modular matrices \cite{KP} for affine Lie
algebras can be computed from the statistics of the representations
with positive energy of associated local current algebras. 

Furthermore, the matrix entries of $S\stat$ were found \cite[II]{FRS} 
to describe the spectrum of the central observables naturally associated 
with the nontrivial topology of the space $S^1$. These discoveries are 
general structure theorems from local quantum field theory and never 
refer to Gibbs functionals (and hardly to conformal invariance). 
They show also, however, that a degeneracy ($S\stat$ being not 
invertible) can -- and in higher dimensions must -- occur which 
obstructs the existence of an $\SL(2,\ZZ)$ representation. 
(Algebraic conditions for non-degeneracy are given in \cite{KLM}.)

Thus, even if $\SL(2,\ZZ)$ does not act on chiral characters, it is
likely to be around, with various caveats as in the 
discussion above, as a consequence of fundamentals of local quantum
field theory, and an interpretation in terms of Gibbs functionals
would be highly desirable. This issue will not be addressed here.

In the classification program for modular invariant 2D partition
functions, it is assumed that certain chiral observables
$A_L\simeq A_R$ are a priori given along with a collection of 
representations (sectors) described by their chiral characters (Gibbs
functionals for the conformal Hamiltonian $L_0$ and suitable other
quantum numbers such as Cartan charges for current algebras). 
These characters transform linearly under the group
$\SL(2,\ZZ)$ which is essentially generated by the imaginary unit shift
($T$) and the inversion ($S$) of the inverse temperature parameter
$\b/2\pi$. One then looks for bilinear combinations of chiral
characters with positive integer coefficients $Z_{l,r}$ (the coupling
matrix) which are invariant under the simultaneous $\SL(2,\ZZ)$  
transformations for both chiral factors (that is, $Z$ commutes with 
$S$ and $T$). The resulting modular invariant partition functions are 
considered as Gibbs functionals for two-dimensional energy and
momentum operators in a representation of a 2D conformally invariant
quantum field theory. The latter contains the chiral observables along with
additional local 2D fields which are nonlocal in each light-cone
coordinate separately. In this interpretation, the entries of the
coupling matrix $Z$ clearly are the multiplicities of the sectors of
the chiral algebras within the representation space of the 2D theory. 
E.g., one usually imposes the constraint $Z_{0,0}= 1$ on the coupling 
matrix which ensures this representation to contain a unique vacuum vector.

One of the most important general classification statements \cite{MS}
asserts that every solution can be turned into a permutation matrix induced 
by an ``automorphism of the fusion rules'' with respect to some ``suitably 
extended algebra of chiral observables'' $A\ext_L\simeq A\ext_R$.
Furthermore, it was found \cite{BE} that the non-vanishing diagonal 
entries of the coupling matrix $Z$ (with respect to the initially given chiral
observables) can be characterized in terms of structure data which refer to
the chiral extension $A\subset A\ext$ only. In the case of SU(2),
these two statements yield the A-D-E classification of \cite{CIZ}.
 
In this article, we endeavour a somewhat opposite program. We assume a
local 2D conformally invariant quantum field theory, denoted by $B$, 
to be given in its vacuum representation $\pi^0$ on a Hilbert space $H$. 
Within this theory we identify chiral observables, denoted by $A\max_L$ 
and $A\max_R$, and show that these are the respective relative commutants 
of any initially given chiral observables $A_R$ and $A_L$ within the same 
2D theory (Corollary 2.7). We then study the superselection sectors of 
the maximal chiral observables which are contained in $H$, that is, 
the branching of the irreducible representation $\pi^0$ upon restriction 
to the subalgebra $A\max_L \otimes A\max_R$. We show that the coupling
matrix for the chiral observables $A\max$ is described by an isomorphism 
between the left and right chiral fusion rules (Corollary 3.5), which 
as a side result implies that $A\max$ coincide with $A\ext$ in the 
modular classification statement (Lemma 3.4). 

We just use the laws controlling local extensions of local algebras, 
as established in \cite{LR}. The crucial point is the fact that the
same coupling matrix which describes the vacuum branching (or the
2D partition function), at the same time describes a distinguished DHR
representation of the chiral observables, and an endomorphism of a von 
Neumann algebra of the form $A_L\otimes A_R$ canonically associated with a
subfactor $A_L\otimes A_R\subset B$. The constraints on the coupling 
matrix arise by the latter endomorphism both being canonical and respecting 
the tensor product (these notions are explained in Sect.\ 3).
 
Unlike locality of the chiral observables, locality of the 2D
net is only implicitly exploited and does not yet enter our (outline
of the) classification itself. It is well known that left and right
chiral sectors (charged fields) cannot be freely composed to yield
local 2D fields \cite{EA,Spt,MS}, and a general algebraic condition in
terms of a statistics operator was given in \cite{LR}. The
incorporation of this condition into our present scheme is still awaiting.

As far as these constraints are concerned, very similar arguments also
apply to ``coset models'' in which a tensor product of two commuting 
subtheories is embedded within a given {\em chiral} theory. 
Therefore, the same constraints on the coupling matrix 
also arise for the branching of the vacuum sector of the ambient 
theory upon restriction to the pair of subtheories. 

The paper is organized as follows. Section 2 sets the physical stage
with emphasis on the equivalence of various possible definitions 
of the chiral observables. In Section 3 the decomposition of the 2D
vacuum representation upon restriction to the chiral observables is
analyzed in the light of the general theory described in \cite{LR}.
The central result is a generalization of the ``automorphism of the 
fusion rules'' theorem \cite{MS}. Section 4 discusses the (first) 
implications for the classification problem. 

The central argument in Section 3 is in fact a theorem on the sector
decomposition of the canonical endomorphism of a von Neumann subfactor. 
This theorem, and the associated notion of a normal canonical tensor 
product subfactor, is of its own mathematical interest \cite{SF} and 
constitutes the common link between various problems in quantum field 
theory, such as chiral observables in 2D, and coset models \cite{X2}
and Jones-Wassermann subfactors \cite{X1,KLM} in chiral conformal 
quantum field theory. Its mathematical essence seems to be most 
appropriately formulated in terms of C* tensor categories. 
It furthermore reveals a connection to asymptotic subfactors \cite{O} 
and quantum doubles \cite{KLM}. This observation may support the 
expected role of quantum double symmetry in 2D conformal quantum 
field theory and coset models.

\section{Chiral observables}

We start with the discussion of various alternatives to define chiral
observables within a conformally invariant 2D theory. The reader
mainly interested in modular invariants is invited to skip this
section, and take its results referred to in Sect.\ 3 for granted.  

We adopt the algebraic approach to quantum field theory in which the 
local algebras are considered rather than the local (Wightman) fields which 
possibly generate them. The underlying picture \cite{LQP} is that the 
{\em net of algebras}, i.e., the complete collection of inclusion and 
intersection relations between algebras associated with smaller and 
larger space-time regions, is sufficient in principle to reconstruct 
the full physical content of the theory. Specifications of the model, 
therefore, have to be formulated as properties of the net of local algebras.

A two-dimensional local conformal quantum field theory is defined on a
covering manifold $\CM$ of Minkowski space-time $\MM=\RR^{(1,1)}$. 
This manifold is obtained as follows \cite{LM,BGL}. One first considers 
Minkowski space-time as the Cartesian product $\RR \times \RR$ of its
two chiral light-cone directions. On each light-cone, 
the M\"obius group $\PSL(2,\RR)$ acts by the rational transformations 
$x\mapsto\frac{ax+b}{cx+d}$, thus enforcing the compactification of 
$\RR$ to $S^1$ by addition of the point $\infty=-\infty$. 
In the quantum field theory, the chiral M\"obius groups are only 
projectively represented, leading to a covering of $S^1$ (in which
$\RR$ will be henceforth identified with the interval $(0,2\pi)$). 
The covering manifold $\CM$ is the Cartesian product of the coverings 
of the two chiral $S^1$, quotiented by the identification 
$(x_L,x_R)= (x_L+2\pi,x_R-2\pi)$. Each subset $(a,a+2\pi)\times(b,b+2\pi)$ 
represents one copy of Minkowski space-time $\MM$ within $\CM$. 

The covering manifold 
$\CM$ possesses a global causal structure such that the causal complement 
of a double cone $O=(a,b)\times(c,d)$\footnote{It is always understood
  that $0<b-a<2\pi$ and $0<d-c<2\pi$.} is the double cone
$O'=(b,a+2\pi)\times(d-2\pi,c)\equiv(b-2\pi,a)\times(d,c+2\pi)$, and
$(O')'=O$. 

We may assume that the 2D theory $B$ is given by the isotonous 
net of local von Neumann algebras $B(O)$ associated with double cones in
Minkowski space-time $O=I\times J\subset\MM$ where $I\subset\RR$ and
$J\subset\RR$ are open intervals on the respective chiral light-cones. 
We assume that $B$ is irreducibly represented on a vacuum Hilbert
space $H$, and transforms covariantly under a strongly continuous
positive-energy representation $U$ of the 2D conformal group. The
latter is the Cartesian product of left and right chiral covering
groups $\widetilde G_L$, $\widetilde G_R$ (with covering projection
$p:\tilde g\mapsto g$) where $G=\PSL(2,\RR)$ is the M\"obius
group. Both chiral M\"obius groups $G$ contain a subgroup U(1) with
positive generators $L_0$, the chiral ``conformal Hamiltonians''.  The
corresponding chiral ``rotations by $2\pi$'' will be denoted for
simplicity by $U_L(2\pi)$ and $U_R(2\pi)$. In a local theory,
$U_L(2\pi)=U_R(2\pi)$, that is, the diagonal of the kernel of 
the covering projection $p$ is represented trivially \cite{LM}.

Conformal covariance means 
$$B(g_LI\times g_RJ) = {\rm Ad}_{U(\tilde g_L,\tilde g_R)} \; B(I\times J)$$
whenever the elements $\tilde g\in \widetilde G$ are represented by 
paths $g_t\in G$ connecting $g$ with the identity which map the
respective chiral intervals pointwise into intervals. If, on the other
hand, the image of an interval under $g_t$ contains $\infty$, then the
above transformation law is considered as the definition of the
algebra on the left hand side, where now  $g_LI$ and $g_RJ$ are
intervals on the covering of the compactified light-cones 
$S^1 = \RR\cup\{\infty\}$. If we denote by $I+2\pi$ and $J+2\pi$ the
images under chiral rotations by $2\pi$, then it follows that
$$B((I+2\pi)\times(J-2\pi))=B(I\times J),$$
that is, the theory $B$ is indeed defined over the conformal covering space
$\CM$. 

Locality of $B$ on Minkowski space implies that the local algebras also 
commute whenever the associated double cones in the covering manifold 
are spacelike separated, i.e., $B(O')\subset B(O)'$. In theories
generated by Wightman fields, one even has

{\bf Essential duality (duality on the covering space $\CM$):} 
$$B(O)'=B(O').$$
The same also holds in parity invariant conformal nets \cite{BGL}. We
shall assume essential duality throughout.

Note that any pair $O$ and $O'$ are a left and a right wedge, or likewise 
the other way round, in a suitable copy of Minkowski space-time in $\CM$,
or can be mapped by M\"obius transformations into these
wedges in any reference copy of Minkowski space-time. Hence, essential
duality is equivalent to wedge duality in Minkowski space. 

We reserve the term Haag duality, according to its original usage \cite{LQP}, 
for the stronger property of duality on Minkowski space $\MM$ (see below) 
which is not an automatic feature. It will not be assumed in this paper.

We proceed to define chiral observables. 

{\bf 2.1.\ Definition: \sl The (maximal) left chiral observables are
$$A\max_L(I):=B(I\times J) \cap U(\widetilde G_R)'.$$ 
The (maximal) right chiral observables $A\max_R(J)$ are defined analogously.}

First we note that this definition does not depend on the interval $J$
since any two open intervals are connected by a M\"obius transformation 
in $\{e\}\times \widetilde G$ which act trivially on $A\max_L(I)$ 
by definition. 
Second, the left chiral observables commute with $U_L(2\pi)=U_R(2\pi)$.
Consequently, the chiral observables are defined over the compactified
light-cone $S^1$ without covering, and are covariant under the proper 
M\"obius group $G=\PSL(2,\RR)$. 
The operators $U_L(2\pi)=U_R(2\pi)$ are multiples
of unity in every irreducible subrepresentation of the chiral 
observables, contained in $H$. The chiral net of von Neumann algebras 
$I\mapsto A\max_L(I)$ satisfies chiral locality (commutativity for 
disjoint intervals) since for given disjoint $I_1$ and $I_2$ it 
is always possible to find intervals $J_1$ and $J_2$ such that 
$O_i=I_i\times J_i$ are space-like to each other, and 2D space-like 
locality of the net $B$ applies. 

Left chiral observables and right chiral observables commute with each
other irrespective of their localization since for any $I$ and $J$ there 
are $\hat J$ and $\hat I$ such that $I\times \hat J$ and $\hat I\times J$ 
are space-like, and again space-like commutativity of $B$ applies.

Clearly, the net $A\max_L$ is M\"obius covariant under the 
representation $U_L\equiv U\vert_{\widetilde G_L}$. By the 
Reeh-Schlieder theorem, the projections $E_L$ onto the subspaces 
$\overline{A_L(I)\Omega}$ for any covariant net $A_L$ do not depend 
on the interval $I$. By standard arguments, involving the Tomita-Takesaki 
modular theory \cite{TT} and exploiting the geometric action of 
the modular group associated with conformal double cone algebras 
\cite{BGL}, one has

{\bf 2.2.\ Lemma: \sl The projection $E_L$ implements a faithful normal 
conditional expectation $\eps_L:B(I\times J)\to A_L(I)$, that is, for 
$b\in B(I\times J)$ there is a unique $a=:\eps_L(b)\in A(I)$ such that
$$E_LbE_L=aE_L.$$
The expectation $\eps_L$ preserves the vacuum state, and the vacuum 
representation of 
the net $A_L$ is given by
$$\pi^L_0(A_L(I))=E_LB(I\times J)E_L.$$
The corresponding statements hold for $A_R$.}

Furthermore, for any M\"obius covariant chiral net, the local algebras, 
unless trivial, are known to be type III von Neumann factors, and one has
\cite{FG2,BGL}

{\bf Essential duality (duality on $S^1$): \sl 
$$\pi_0(A(I))'=\pi_0(A(I'))$$
valid in the vacuum representation $\pi_0$ of $A$.}

Hence the chiral observables automatically satisfy essential duality.

{\bf 2.3.\ Lemma: \sl The subspace $\overline{A\max_L(I)\Omega}$ 
coincides with the subspace of $U_R$-invariant vectors in $H$, that is
$$E'_L=E\max_L$$
where $E'_L$ denotes the projection onto the $U_R$-invariant
subspace. The corresponding statement holds for $A_R$.} 
 
{\em Proof:} I owe the following argument to D. Buchholz. We only have
to show that every $U_R$-invariant vector can be approximated in
$A\max_L(I)\Omega$. 
Since $B(O)\Omega$ is dense in $H$, $E'_LB(O)\Omega$ is dense
in $E'_LH$. Consider any vector $\Psi=E'_Lb\Omega$ with $b\in B(O)$. 
Then $\Psi=U_R(g)\Psi=E'_L\a_g(b)\Omega$ for all $g\in\widetilde G_R$, and 
$\Psi=E'_Lb_T\Omega$ where 
$$b_T=\frac 1{2T}\int_{-T}^Tdt\,\a_{g_t}(b)$$ 
is an average over the one-parameter group of right chiral dilatations
$g_t$ which leave the interval $J$ fixed. Since 
$\Vert b_T\Vert\leq\Vert b\Vert$, the family $b_T$ has a weak limit 
point $a$ in the von Neumann algebra $B(O)$ as $T\to\infty$. We are 
going to show that $a$ is invariant under $\widetilde G_R$, hence 
commutes with $U_R$ and thus belongs to $A\max_L(I)$. It follows that
$\Psi=E'_Lb_T\Omega=E'_La\Omega=a\Omega$ is in $A\max_L(I)\Omega$, and
the latter space is dense in $E'_LH$. 

In order to show the $\widetilde G_R$-invariance of $a$, we first note that
$$\Vert\a_{g_s}(b_T)-b_T\Vert=\frac 1{2T}\Vert\left[\int_{-T+s}^{-T} +
\int_{T}^{T+s}\right]dt\, \a_{g_t}(b)\Vert\leq\frac{\vert s\vert}T\Vert b
\Vert$$ 
which vanishes as $T\to\infty$. Hence $a$ is dilatation invariant, and 
$$\Vert\a_g(a)-a\Vert=\Vert\a_g\a_{g_t}(a)-\a_{g_t}(a)\Vert=
\Vert\a_{g_{-t}gg_{t}}(a)-a\Vert$$ 
for all $g\in \widetilde G_R$ and all $t$. For $g$ a translation resp.\ 
a special conformal transformation (relative to the dilatations $g_t$), 
$g_{-t}gg_{t}$ tends to the identity as $t\to-\infty$ resp.\ $t\to+\infty$. 
For $a$ sufficiently regular to have norm-continuity of $\a_g$ (which
is the case if $b$ above was regular; such operators still generate a
dense subspace of $H$) it follows that $\Vert\a_g(a)-a\Vert=0$, as asserted.
\qed  

We want to study the equivalence of the Definition 2.1 with
several alternative reasonable definitions. For this purpose,
we first compile some useful notions for 2D and for chiral nets. 

{\bf Generating property:} The net $B$ is said to have the 
{\rm generating property} if
$$U(\widetilde G_L)\subset B(I\times J)\vee B(I'\times J)$$
for any $J$, and equivalently (taking commutants and using essential 
duality of $B$) if
$$B(I\times J)\cap B(I'\times J)\subset U(\widetilde G_L)'$$
for any $J$. (Here $I'$ is either of the two intervals $I^+=(b,a+2\pi)$ 
or $I^-=(b-2\pi,a)$ if $I=(a,b)$. By the very second formula, the
algebra on its left hand side does not depend on this choice, since a
suitable left M\"obius transformation which maps $I$ onto $I^+$ and $I^-$ 
onto $I$, leaves the intersection of the two algebras invariant.) 

{\bf Haag duality (duality on Minkowski space $\MM$):} The net $B$
fulfils {\rm Haag duality} if 
$$B(O)'=B(O^c)\equiv B(O^-)\vee B(O^+)$$
where $O^c$ is the disconnected causal complement of $O$ in
Minkowski space with connected components $O^-$, $O^+$.

{\bf Strong additivity:} The net $B$ fulfils {\rm strong additivity} if
$$B(O_1)\vee B(O_2)=B(O)$$
for $O_1$ and $O_2$ the two connected components of the causal
complement of an interior point in a double cone $O$. 

{\bf Chiral additivity:} The net $B$ fulfils {\rm chiral additivity} if
$$B(I_1\times J)\vee B(I_2\times J)=B(I\times J)$$
if $I_1$, $I_2$ arise from $I$ by removal of an interior point; and
likewise for the two light-cone directions interchanged.

{\bf Generating property:} A left chiral net $A_L$ of subalgebras of 
$A\max_L(I)$ has the {\rm generating property} if 
$$U(\widetilde G_L)\subset A_L(I)\vee A_L(I');$$
the analogous definition holds for right chiral nets $A_R$ of
subalgebras of $A\max_R(J)$. 

{\bf Haag duality (duality on $\RR$):} A chiral net $A$ fulfils {\rm
  Haag duality} if 
$$\pi_0(A(I))'=\pi_0(A(I^c))$$
holds in the vacuum representation. Here $I^c$ denotes the 
(disconnected) open complement of an interval $I$ in $\RR$. 

{\bf Strong additivity:} A chiral net $A$ fulfils {\rm strong additivity} if
$$A(I_1)\vee A(I_2)=A(I)$$ 
if $I$ is an interval in $S^1$ divided into two subintervals 
$I_1$, $I_2$ by removal of an interior point. 

It is obvious, that if any net $A_L$ of subalgebras of $A\max_L$ has
the generating property, then $A\max_L$ has the generating property
and $B$ also has the generating property.

In fact, in view of the previous definition of chiral observables, 
the generating property for $A\max$ is actually a feature of the 2D net $B$.
In the cyclic subspace of the chiral observables (their vacuum
representation), the assumption is always true by essential duality and 
factoriality. But the generating property for $A$ 
is required to hold on the full
vacuum Hilbert space $H$ of $B$. It certainly holds if $B$ possesses a
conserved stress-energy tensor whose chiral components then are among 
the chiral observables. It also holds, e.g., in the theory generated by the 
derivatives of a massless conserved vector current which has nontrivial 
chiral observables (the derivatives of a U(1) current)
but no stress-energy tensor. Namely, in this model, 
$B(I\times J)= A\max_L(I)\otimes A\max_R(J)$, and $H=H_L\otimes H_R$. 
Thus $H$ contains only the vacuum representation of the chiral observables. 
Therefore, we believe that the assumption of the generating property 
for chiral observables does not exclude any models of serious interest. 

The following assertions hardly need to be proven.

{\bf 2.4.\ Lemma: 
\rm(i) \sl Haag duality is equivalent to strong additivity, both for
2D and chiral conformal nets. 

\rm(ii) \sl Strong additivity of a 2D net implies chiral additivity.}

{\em Proof:} (i) By essential duality, Haag duality is equivalent to 
$B(O^c)=B(O')$ (in the 2D case). This in turn is strong additivity
since, in the covering space $\CM$, the two connected components of
$O^c$ touch each other in a point (``space-like infinity''), and thus 
constitute the causal complement of that point in $O'$. The same
argument applies in the chiral case, as the two connected
components of $I^c$ touch each other in $S^1$ at infinity. \\
(ii) Let $J_1$ and $J_2$ arise by removal of an arbitrary
interior point from $J$, such that $O_1=I_1\times J_2$ and
$O_2=I_2\times J_1$ are the components of the causal complement of
an interior point in the double cone $O=I\times J$. Then $B(O_1)\vee
B(O_2)\subset B(I_1\times J)\vee B(I_2\times J)\subset B(O)$, and
strong additivity implies equality. \qed

In order to compare alternative definitions of chiral observables, we
consider the following two chains of inclusions, which hold just by
isotony and essential duality: 

{\bf 2.5.\ Lemma: \sl With notations as explained below, one has}
\begin{eqnarray*} A\max_L(I_2)\;\subset\; \bigcap_J B(I_2\times J)\;\subset\;
B_{2,\hat 1}\cap B_{2,\hat 2} \;\subset\; B_{2,1}\cap B_{2,2} \;\subset\;
\quad\qquad  \\ \left\{\begin{array}{l}\;\subset\; 
B_{2+3,1}\cap B_{2,2}\;\equiv\; B_{1,2}'\cap B_{2,2} \;\subset\; 
A_R(J_2)'\cap B_{2,2}\;. \\ \\ \;\subset\; B_{2+3,1}\cap B_{1+2,2}\;\equiv\; 
B_{1,2}'\cap B_{1+2,2}\;\equiv\; B_{2+3,1}\cap B_{3,1}'\;.\end{array} \right. 
\end{eqnarray*}
Here we have picked three left chiral intervals $I_1=(0,a)$,
$I_2=(a,b)$, $I_3=(b,2\pi)$ and two right chiral intervals
$J_1=(0,c)$, $J_2=(c,2\pi)$ as indicated by the figure, and employ 
short hand notations $B_{i,j}=B(I_i\times J_j)$, 
$B_{i,\hat j}=B(I_i\times \hat J_j)$ with $\hat J_j\subset J_j$. The 
labels $1+2$ resp.\ $2+3$ stand for the intervals $(0,b)$ resp.\ $(a,2\pi)$. 

\begin{center}\begin{picture}(130,120)
\multiput(70,5)(-20,20){4}{\line(1,1){60}}
\multiput(70,5)(30,30){3}{\line(-1,1){60}}
\put(63,0){\vector(-1,1){59}} \put(77,0){\vector(1,1){59}}
\put(44,19){\circle*{1}} \put(24,39){\circle*{1}} \put(106,29){\circle*{1}}
\put(46,1){$I_1$} \put(26,21){$I_2$} \put(6,41){$I_3$}
\put(90,8){$J_1$} \put(120,38){$J_2$}
\put(69,27){\small 1,1} \put(49,47){\small 2,1} \put(29,67){\small 3,1} 
\put(99,57){\small 1,2} \put(79,77){\small 2,2} \put(59,97){\small 3,2} 
\put(67.3,-5){$0$} \put(131,63){$2\pi$} \put(-3,63){$2\pi$}
\put(44,20){$a$} \put(24,40){$b$} \put(100,30){$c$}
\put(130,90){$(0,2\pi)\times (0,2\pi)=\MM$}
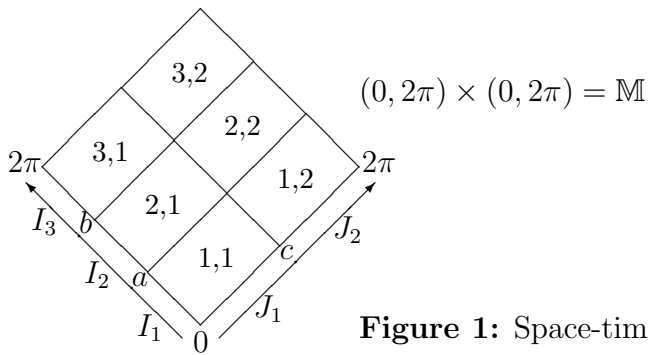
\end{picture}{\bf Figure 1:} Space-time regions in Lemma 2.5\end{center}

Of course, the choice of the values $0<a<b<2\pi$ and $0<c<2\pi$ is 
completely immaterial since the ensuing partition of one copy of Minkowski
space $\MM$ within the covering space $\CM$ can be transferred to any
other partition of any 
other copy by left and right M\"obius transformations.

$A_R$ in the second line in Lemma 2.5 is any covariant net of
subalgebras of $A\max_R(J)$. The consideration of subalgebras of the
maximal chiral observables is motivated by our intention to compare
with the context of modular invariant partition functions. There one
usually starts with some a priori given chiral observables $A_R$ and
$A_L$ such as current algebras while the maximal ones might turn out
as some ``$W$-algebra'' extension thereof. Indeed, we shall later find
a condition (Corollary 3.5) when the given chiral observables and the
maximal ones coincide. 
 
Of particular interest are the expressions $\bigcap_J B(I_2\times J)$,
$B_{1,2}'\cap B_{1+2,2}$, and $A_R(J_2)'\cap B_{2,2}$ figuring in
Lemma 2.5. The first one is possibly nontrivial even in massive 2D
theories \cite{SW}, where it provides a ``holographic'' sattelite 
theory (with a conformal symmetry emerging automatically \cite{HWW});
it has been used as a definition of observables on a horizon in curved
space-time \cite{GLRV} in the absence of space-time symmetries.
The second one is, up to a M\"obius transformation, the relative commutant 
of a wedge algebra $B(W+a)$ within another wedge algebra $B(W)$ 
where $a$ is a shift in a light-like direction. The third one is the 
relative commutant of the opposite chiral observables within a double 
cone. Each of these would be a sensible definition of chiral observables.

In fact, under suitable conditions, the inclusions above turn into 
equalities and the various definitions coalesce. (Note that any
nontrivial inclusion in the first and second line would require the 
respective larger algebra not to commute with $U(\widetilde G_R)$.)

{\bf 2.6.\ Proposition \sl (referring to the chains of inclusions as in
  Lemma 2.5):

\rm(i) \sl If $B$ has the generating property, then all
inclusions in the first line are equalities.\\
\rm(ii) \sl If $A_R$ has the generating property, then all
inclusions in the first and second lines are equalities.\\
\rm(iii) \sl If all inclusions in the first and second line are
equalities, then $B$ has the generating property.\\
\rm(iv) \sl If $B$ satisfies Haag duality, then $B_{2,\hat 1}\cap
B_{2,\hat 2}=B_{1,2}'\cap B_{1+2,2}$ where $\hat J_1=(c',c), \hat
J_2=(c,c'')$, $0<c'<c<c''<2\pi$; in particular, if in addition the 
inclusions in the first line are equalities, then all inclusions in
the first and third lines are equalities.\\
\rm(v) \sl If all inclusions in the first line are equalities and $B$
satisfies chiral additivity, then $A\max_L$ satisfies Haag duality. 
The corresponding statement holds for $A\max_R$.}

Versions of assertions (iv) and (v) are also contained in \cite{GLRV}.

{\it Proof:} (i) The generating property of $B$ implies 
that $B(I\times J) \cap B(I\times J')$ commutes with 
$U(\widetilde G_R)$ and thus is contained in $A\max_L(I)$.\\
(ii) $B_{2,2}$ commutes with $A_R(J_1)$, and hence $A_R(J_2)'\cap B_{2,2}$ 
is contained in $[A_R(J_2)'\cap A_R(J_1)']\cap B_{2,2}$ which by the 
generating property of $A_R$ is contained in $U(\widetilde G_R)'\cap B_{2,2}
=A\max_L(I_2)$. \\
(iii) We have $U(\widetilde G_R) \subset A\max_L(I_2)' = B_{1,2}\vee
B_{3,1}$ by definition and  by assumption. The claim follows by isotony
with $I_1\times J_2 \subset (b-2\pi,a)\times J_2$ and $I_3\times J_1 
= (I_3-2\pi)\times (J_1+2\pi) \subset (b-2\pi,a)\times J_2$. \\
(iv) We have $B_{1,2}'\cap B_{1+2,2}=B_{1,2}'\cap B_{3,1}'$. By isotony, 
this is contained in $B_{1,2}'\cap B_{3,\check 1}'$ which equals 
$B_{2,\hat 1}$ by Haag duality, where $\check J_1=(0,c')$. 
The same algebra is similarly 
contained in $B_{1,\check 2}'\cap B_{3,1}'=B_{2,\hat 2}$ where 
$\check J_2=(c'',2\pi)$. This gives both assertions, 
by inspection of the chain of inclusions, Lemma 2.5. \\
(v) Chiral additivity for $B$ is, by passing to the commutants, 
equivalent to 
$$B((a,b)\times J)=B((0,b)\times J)\cap B((a,2\pi)\times J)$$
for $0<a<b<2\pi$ and any interval $J$. Taking suitable intersections over 
$J$ to obtain $A\max_L(I)$ by using equality in the first line of the 
chain of inclusions, yields
$$A\max_L((a,b))=A\max_L((0,b))\cap A\max_L((a,2\pi)).$$
Since the vacuum representation is faithful on $\RR\cong(0,2\pi)$, the same
holds in the vacuum representation $\pi_0$ (i.e., after multiplication with 
$E\max_L$). Passing to the commutants in the vacuum representation, 
and using essential duality for $A\max_L$, one gets strong additivity 
in the vacuum and hence in any representation. 
\qed

We must admit a little lapse in the proof of (v). Namely, the vacuum 
representation of a chiral net $A$ is known to be faithful on the 
quasilocal C* algebra on
$\RR\cong(0,2\pi)$ which does not contain the von Neumann algebras 
$A((0,b))$ and $A((a,2\pi))$. Yet, we are confident that the above 
conclusion from faithfulness is correct for the intersections. We have
tested its validity in the (prototypical) model with a chiral U(1)
current $j$ and associated charge $Q=\int j(x)dx$. The operator
$\exp itQ$ which is trivially represented in the vacuum representation
can be weakly approximated by Weyl operators $\exp itj(f_R)$ as
$R\to\infty$ where $f_R(x)=f(x/R)$ and $f$ is a testfunction with
$f(x)=1$ for $\vert x\vert<1$ and $f(x)=0$ for $\vert x\vert>2$, say. 
Splitting $f_R$ into two pieces $f_R^\pm$ with supports in $(-2R,a)$ and in
$(-a,2R)$ respectively, yields two Weyl operators $\exp itj(f_R^-)$
and $\exp itj(-f_R^+)$ localized in overlapping left and right halfspaces 
whose weak limits as $R\to\infty$ should coincide in the vacuum 
representation, and differ in a charged representation by a factor 
$\exp itQ$. Nontriviality of these weak limits would invalidate our 
conclusion in the proof of (v). A calculation, however, shows that, 
due to scale invariance, the cutoff within the fixed interval $(-a,a)$
in comparison to the increase in $R$ behaves like a cutoff in a scaled
interval $(-a/R,a/R)$, and produces an ultraviolet singularity which 
causes the weak limits of interest to be zero. Since this ultraviolet 
behaviour is a ``universal'' effect of scale invariance, we believe
that the same mechanism protects the validity of our conclusion also 
in general models. In any case, (v) will not be needed for the
purposes of this paper. 

Since we consider the assumption of the generating property for the
chiral observables as no serious restriction, we reformulate the
statements with this assumption as a default.

{\bf 2.7.\ Corollary: \rm (i) \sl Assume the generating property for some 
  nets $A_L(I)$ and $A_R(J)$ of subalgebras of $B(I\times J)$ which are
  invariant under the respective opposite M\"obius group. Then
$$A\max_L(I)=\bigcap_J B(I\times J)=A_R(J)'\cap B(I\times J),$$
  and similarly for $A\max_R$. In particular, the left and right maximal
  chiral observables are each other's mutual relative commutants in $B$. 

\rm (ii) \sl If the net $B$ is Haag dual, then $A\max_L$ and $A\max_R$
  are Haag dual, and 
$$A\max_L(I_1)=B(I_2\times J)'\cap B(I\times J)$$
  where $I_1$, $I_2$ arise from the interval $I$ by removal of an 
  interior point, and $J$ is an arbitrary interval.
  The corresponding statement holds for $A\max_R$.} 

(Again, the assertion of Haag duality for the chiral observables has to
be taken with a little caution.)

We conclude this section with a study of the joint position of the 
subalgebras of left and right chiral observables within $B(O)$. We have

{\bf 2.8.\ Proposition: \sl In the vacuum representation of $B$, the left
  and right chiral observables are in a tensor product position, i.e.,}
  $$A\max_L(I)\vee A\max_R(J)\simeq A\max_L(I)\otimes A\max_R(J).$$ 

{\it Proof:} The statement follows, by Tomita-Takesaki modular theory
\cite{TT}, from the existence of the conditional expectations $\eps_L$ and
$\eps_R$, cf.\ Lemma 2.2. We want to give a less abstract argument. 

Since left and right chiral observables mutually commute,
it is sufficient to consider products $a_La_R$ where $a_L\in A\max_L(I)$
and $a_R\in A\max_R(J)$. Since the vacuum state $\omega$ is conformally 
invariant, and since the chiral observables transform under the respective 
chiral M\"obius groups only, we have
\begin{eqnarray*} \omega(a_La_R)=\omega(\a_{g_L\times g_R}
(a_La_R))= \omega(\a_{g_L}(a_L)\a_{g_R}(a_R)).\end{eqnarray*}
For suitable elements $g_L$ and $g_R$, the localization of the
transformed observables tends to space-like infinite separation, hence
the cluster property of the vacuum state applies and entails
$$\omega(a_La_R)=\omega(a_L)\omega(a_R).$$
The factorization of the (normal) vacuum state implies the tensor product
factorization of the corresponding algebras. \qed

\section{Representation theory}
A subtheory $A$ of a given theory $B$ is described by a net of subalgebras
(subfactors) $A(O)\subset B(O)$. Conversely, $B$ may be considered as
a (local) extension of a given theory $A$. In the present paper, 
$A$ is a net of left and right chiral 
observables\footnote{Henceforth, the notation $O=I\times J$ 
   will be understood.} 
$O\mapsto A(O)=A_L(I)\otimes A_R(J)$, contained in a two-dimensional
net $O\mapsto B(O)$. 

A general analysis of the representation theory in this situation 
was initiated in \cite{LR}. As a prerequisite it was required that, 
in generalization of an unbroken global gauge symmetry, there is a 
consistent family of (normal, faithful) conditional expectations 
$\eps_O:B(O)\to A(O)$ which commute with space-time symmetries and 
preserve the vacuum state.
 
In our situation at hand, these expectations are provided by 
Takesaki's theorem \cite{TT}, thanks to the fact that Tomita's modular
group for conformal double cone algebras is a subgroup of 
$\widetilde G_L\times \widetilde G_R$ and consequently preserves any 
M\"obius covariant subtheory of the form $A_L\otimes A_R$. As in
Sect.\ 2, they are coherently implemented by the projection $E_{LR}$ onto 
the closure of the 
subspace $A_L(I)A_R(J)\Omega$ (not depending on $I\times J$), which  
commutes with M\"obius transformations and preserves the vacuum state. 

Actually, for the analysis in \cite{LR} nets have to be {\em directed}. 
We must therefore pass to the 2D and chiral theories on Minkowski 
space $\MM$ and the light-cone axes $\RR$, respectively. As is common 
practice, we denote the quasilocal C* algebra generated by a directed 
net of von Neumann algebras (say $A(O)$) by the same symbol (say $A$) 
as the net itself. We also denote the vacuum representations of $A$ 
and of $B$ by $\pi_0$ and $\pi^0$, respectively.

In the algebraic approach to quantum field theory, positive energy
representations are conveniently described in terms of DHR
endomorphisms \cite{LQP}, provided Haag duality holds. But the 
restriction of $\pi^0$ to the subtheory $A$ is always given by a DHR 
endomorphism $\rho$ of $A$ 
$$\pi^0\vert_A\simeq\pi_0\comp\rho$$
even without assuming Haag duality \cite{LR}. Moreover, $\rho$ is of the
``canonical'' form $\rho=\bar\iota\comp\iota$. Here $\iota:A\to B$ 
is the embedding homomorphism and $\bar\iota:B\to A$ is a conjugate
homomorphism to $\iota$ in the sense \cite{Dim} that there exist isometric
intertwiners $w\in A, w:\id_A\to\bar\iota\comp\iota\equiv\rho$ and 
$v\in B, v:\id_B\to\iota\comp\bar\iota\equiv\gamma$ with 
$w^*v=w^*\g(v)=\lambda\inv\cdot\eins$. The number $\lambda \geq 1$ is the 
(statistical) dimension of $\rho$ and coincides with the index of the 
local subfactor $A(O)\subset B(O)$ which is independent of $O$.
(We assume this index, and hence the dimensions of $\rho$ and all its
subsectors, to be finite throughout.)

The construction given in \cite{LR} starts off from a canonical 
endomorphism \cite{L} $\g_O$ of the local von Neumann algebra 
$B(O)$ for any fixed double cone $O$ into its subfactor $A(O)$.
$\g_O$ extends to a canonical endomorphism $\g$ of the quasilocal algebra 
$B$ into $A$ in such a way that on any $B(\hat O)$, $\hat O\supset O$, it 
yields a canonical endomorphism of $B(\hat O)$ into $A(\hat O)$, and
consequently the restriction of $\rho=\g\vert_A$ to $A(\hat O)$ is the 
corresponding dual canonical endomorphism. It was shown that $\rho$ is
a DHR endomorphism localized in the fixed double cone $O$, and that
$w\in A(O)$ and $v\in B(O)$ are local operators. 

In the present case, $A$ being a tensor product $A_L\otimes A_R$ of C* 
algebras, any irreducible representation is also a C* tensor product. As
pointed out by R. Longo, there is a theoretical possibility (in case
the chiral representations are not ``type I'', cf.\ \cite{KLM}), that
the C* tensor products are not spatial. In a large class of models,
including current algebras, this possibility can be ruled out
\cite[Lemma 12]{KLM}, however, and it can presumably never arise when
the statistical dimension is finite. Thus we may assume that the
corresponding subspaces of $H$ are also tensor products.

Let therefore the irreducible decomposition of the restricted vacuum
representation into chiral sectors be given by
$$\pi^0\vert_{A_L\otimes A_R} \simeq \bigoplus_{l,r} Z_{l,r}\;
\pi^L_{l}\otimes \pi^R_{r}$$
with a (possibly rectangular) matrix of nonnegative integers $Z_{l,r}$
where $l,r$ run over the irreducible superselection sectors of the
left and right chiral observables contained in $H$. Equivalently, the
corresponding DHR endomorphism $\rho$ decomposes as
$$\rho\simeq\bigoplus_{l,r} Z_{l,r}\;\rho^L_{l}\otimes\rho^R_{r}$$
with irreducible chiral DHR endomorphisms $\rho^L_{l}$ and $\rho^R_{r}$, 
and with the same matrix $Z$. We call $Z$ the {\bf coupling matrix},
and we reserve the labels $l=0$ and $r=0$ for the respective vacuum sectors, 
$\rho^L_{0}\simeq\id_{A_L}\equiv\id_L$ and 
$\rho^R_{0}\simeq\id_{A_R}\equiv\id_R$.

Making contact with modular invariants, it should be clear that 
the coupling matrix also enters the decomposition of the vacuum 
partition function of a 2D local theory 
$$\hbox{Tr}_{\pi^0} \;e^{-\beta (L^L_0+L^R_0)} =
\sum_{l,r} Z_{l,r}\;\hbox{Tr}_{\pi^L_l} \;e^{-\beta L^L_0}\;
\hbox{Tr}_{\pi^R_r} \;e^{-\beta L^R_0}  $$ 
into chiral characters $\chi_\pi=\hbox{Tr}_{\pi}
\;e^{-\beta L_0}$ of the representations of the chiral observables.

A similar algebraic situation with a tensor product of two nets of
observables embedded into another net also arises in 
coset models \cite{X2} in chiral quantum field theory.
These models are given by a net of chiral observables $B(I)$ and a
proper subnet $A(I)$ (e.g., the current algebras associated with a 
compact Lie group $G$ and a subgroup $H$). The coset theory is defined
as the net of relative commutants $C(I):=A(I)'\cap B(I)$. 
Unless the pair of groups gives rise to a conformal inclusion (in
which case $C(I)$ is trivial), the net 
$C$ possesses a stress-energy tensor of its own which 
commutes with the stress-energy tensor of $A$. An argument similar
as in Proposition 2.8, making use of the two commuting M\"obius groups 
for $A$ and $C$, yields the tensor product position of $A$ and $C$
within $B$. Again, the branching of the vacuum sector of $B$ is
described by a coupling matrix, and our results below can be easily
adapted to coset models.

We are going to study the branching of the vacuum representation
$\pi^0\vert_A$ in terms of the endomorphism $\rho$. It turns
out convenient to do this in a framework of endomorphisms of von
Neumann algebras. For this purpose we use the fact that $\rho$ as a DHR
endomorphism of the quasilocal algebra $A$ has the same decomposition
into irreducibles as its restriction $\rho_O=\rho\vert_{A(O)}$ as a 
(dual canonical) endomorphism of a local von Neumann algebra. This
statement is standard if one assumes Haag duality and strong additivity. 
But it has also been established without these assumptions in the
chiral case, making use of conformal symmetry and essential duality
instead, provided the statistical dimension is finite \cite{GL}. The
latter argument carries over without difficulty to the 2D case.  
We just state this result without repeating its proof. 

{\bf 3.1.\ Lemma: \sl Let $A$ be a local net on $\MM$ or $\RR$. Assume
  either that $A$ is the restriction of a conformal net on $\CM$
  resp.\ $S^1$, or that $A$ satisfies Haag duality and
  strong additivity. Let $\sigma$, $\tau$ be two DHR endomorphisms 
  (in the conformal case: with finite statistical dimension),
  localized in some double cone or interval $O$, and $\sigma_O$, $\tau_O$ their restrictions to
  $A(O)$. Then the intertwiner spaces $(\sigma,\tau)$ and
  $(\sigma_O,\tau_O)$ coincide. In particular, $\sigma$ and $\sigma_O$
  have the same decomposition into irreducibles.} 

Since our nets $B$ and $A_L$, $A_R$ are conformal, the Lemma applies
to all DHR endomorphisms with finite dimension. It follows that the 
decomposition
$$\rho_O\simeq\bigoplus_{l,r} Z_{l,r}\;\rho^L_{l}\otimes\rho^R_{r}$$
of the dual canonical endomorphism for the local subfactor
$A_L(I)\otimes A_R(J)\subset B(O)$ is again described by the same
coupling matrix $Z$, where now $\rho^L_l$ and $\rho^R_r$ are 
local restrictions of chiral DHR endomorphisms. 

The crucial additional information here is that $\rho$ and hence the
dual canonical endomorphism $\rho_O$ respects the tensor product 
structure $A(O)=A_L(I)\otimes A_R(J)$ in the sense that its
irreducible components are equivalent to tensor products of
irreducible endomorphisms of the factor algebras. We call a 
von Neumann subfactor 
$A\otimes C\subset B$ with this property a {\bf canonical tensor product 
subfactor (CTPS)}\footnote{An elementary example of a
  subfactor $A\otimes C\subset B$ which is {\em not}
  canonical in this sense was suggested to me by H.J.~Borchers: 
  take $C=A$, and $B$ the crossed product of $A\otimes A$ by the 
  flip automorphism. Then the dual canonical endomorphism is
  the direct sum of the identity and the flip. The latter does not
  respect the tensor product.} with associated coupling matrix $Z$.

The subfactors $A_L(I)\otimes A_R(J)\subset B(O)$, or 
$A(I)\otimes C(I)\subset B(I)$ for coset models, are examples of
CTPS's. Other examples in conformal quantum field theory are 
Jones-Wassermann subfactors arising from partitions of $S^1$ into 
four intervals \cite{X1,KLM}. 

Since we assume the index to be finite, only finitely many sectors can
contribute which all must have finite dimension, hence the coupling matrix
is a finite matrix. Since we have assumed the defining representation
of $B$ to contain a unique vacuum vector, it follows that its
restriction to the chiral observables contains the joint vacuum
representation exactly once, hence $Z_{0,0} = 1$. This implies that
the multiplicity of $\id_L\otimes \id_R$ in $\rho$ is one, hence
the embedding $A_L\otimes A_R\subset B$ is irreducible (both for the
local von Neumann algebras and for the quasilocal C* algebras).

We summarize the discussion so far:

{\bf 3.2.\ Proposition: \sl The local subfactors $A_L(I)\otimes A_R(J)
  \subset B(O)$ are irreducible canonical tensor product subfactors. 
  The irreducible sector decomposition of their dual canonical 
  endomorphisms is described by the same finite coupling matrix $Z$ as the 
  decomposition of the restricted vacuum representation
  $\pi^0\vert_{A_L\otimes A_R}$ of $B$.}

We are going to study the constraints on $Z$ being the coupling matrix
of a canonical TPS.
These constraints are then read back as constraints on the representation 
$\pi^0\vert_{A_L\otimes A_R}$ or on the 2D partition function.

In the sequel when we write $A_L\otimes A_R\subset B$, we have in
mind the local subfactor $A_L(I)\otimes A_R(J)\subset B(O)$, or with
suitable modifications $A(I)\otimes C(I)\subset B(I)$ in coset
models. But we are actually going to establish general statements on 
coupling matrices of CTPS's without reference to quantum field theory.

We shall several times need ``Frobenius reciprocity'', cf.\ \cite{Dim}, 
which we recall in 

{\bf 3.3.\ Lemma: \sl Let $A$, $B$, $C$ be unital C* or von Neumann
  algebras and $\a:A\to B$, $\beta:B\to C$, $\g:A\to C$ unital 
  homomorphisms. Denote by $\langle\g,\a\b\rangle$ the dimension of 
  the linear space of intertwiners $t\in C$, $t:\g\to\a\b$. Then 
$$\langle\bar\a\g,\b\rangle=\langle\g,\a\b\rangle=\langle\g\bar\b,\a\rangle$$
provided the conjugate homomorphisms $\bar\a:B\to A$ or 
$\bar\b:C\to B$ exist.}

Here, as before, conjugates are defined in terms of a pair of 
intertwiners \cite{Dim}, say $w:\id_A\to\bar\a\a$, $v:\id_B\to\a\bar\a$ 
which satisfy the relations $\a(w)^*v=\eins_B$, $\bar\a(v)^*w=\eins_A$.

For $X\subset B$ the relative commutant $X'\cap B$ is commonly denoted by
$X^c$. We have

{\bf 3.4.\ Lemma: \sl Let $A_L\otimes A_R\subset B$ be a CTPS with
  finite index, and $Z_{l,r}$ its coupling matrix.
  Then, $Z_{0,r}\neq 0$ implies $r=0$ if and only if 
  $\eins\otimes A_R=(A_L\otimes \eins)^c$.   
  The corresponding statement holds exchanging $A_L$ and $A_R$.} 

{\it Proof:} We have to show that $Z_{0,r}\neq 0$ for some $r\neq 0$ 
(that is, $\rho^R_r\not\simeq\id_R$) if 
and only if the inclusion $\eins\otimes A_R\subset (A_L\otimes \eins)^c$ 
is proper. Note that equality holds if and only if 
$X:=(A_L\otimes\eins)\vee (A_L\otimes\eins)^c$ equals $A_L\otimes A_R$
(since $A_L$ is a factor).

Consider now the intermediate subfactor $A_L\otimes A_R\subset X\subset B$. 
In terms of the inclusion maps
$\iota_1:A_L\otimes A_R\to X$ and $\iota_2:X\to B$, we have 
$$\rho_1\equiv\bar\iota_1\iota_1\prec
\bar\iota_1\bar\iota_2\iota_2\iota_1=\bar\iota\iota\equiv\rho.$$
If $A_L\otimes A_R\subset X$ is proper, then $\iota_1$
is nontrivial and $\rho_1$ contains a nontrivial subsector 
$\id_L\otimes\rho^R_{r}$ which is also a subsector of $\rho$ giving rise
to a nonvanishing matrix entry $Z_{0,r}$. Conversely, if 
$Z_{0,r}\neq 0$, then $\id_L\otimes\rho^R_r\prec\rho$. By Frobenius 
reciprocity (Lemma 3.3), $\iota\prec\iota\comp(\id_L\otimes\rho^R_r)$,
hence there is a nonvanishing intertwiner $\psi\in B$ which satisfies 
$$\psi(a_L\otimes a_R)=(a_L\otimes \rho^R_{r}(a_R))\psi.$$
Putting $a_R=\eins$, this implies that $\psi\in (A_L\otimes\eins)^c$,
thus $\psi\in X$, and hence $\iota_1\prec\iota_1\comp(\id_L\otimes\rho^R_r)$. 
Again invoking Frobenius reciprocity, $\id_L\otimes\rho^R_{r}\prec\rho_1$. 
Thus $A_L\otimes A_R\subset X$ is proper. \qed  

The Lemma allows us to characterize the maximal chiral observables by
a normality property of the local subfactors, see Corollary 3.5 below. 
We recall that an inclusion $A\subset B$ is called normal if
$(A^c)^c=A$. In general, $A^{cc}\supset A$. It follows that
$(A^{cc})^c\subset A^c$ and $(A^c)^{cc}\supset A^c$, hence
$A^{ccc}=A^c$ which is obviously equivalent to the statement that a
relative commutant is always normal.

We call (with a slight abuse of terminology) a tensor product 
subfactor $A\otimes C\subset B$ {\bf normal} if $A\otimes\eins$ and
$\eins\otimes C$ are each other's relative commutants in $B$.
 
Hence, the local subfactors of chiral observables within 2D conformal
quantum field theories, $A\max_L(I)\otimes A\max_R(J)\subset B(O)$ 
are examples of normal and canonical TPS's. 
Also coset models give rise to local subfactors which are normal CTPS's. 
Namely, one obtains normality by extending (if necessary) $A(I)$ by the
relative commutant of $C(I)$.

Normality of the local subfactors is characteristic for the maximal
chiral observables, and a criterium in terms of the coupling matrix is
given in

{\bf 3.5.\ Corollary: \sl The following are equivalent. 

\rm (i) \sl $A_L=A\max_L$ and $A_R=A\max_R$.\\
\rm (ii) \sl The local subfactors $A_L(I)\otimes A_R(J)\subset B(O)$ 
are normal CTPS's.\\
\rm (iii) \sl The coupling matrix satisfies $Z_{0,r}=\delta_{0,r}$ and  
  $Z_{l,0}=\delta_{l,0}$.\\
\rm (iv) \sl The coupling matrix describes an isomorphism of the left and
right chiral fusion rules (in the sense of Theorem 3.6 below).}

{\it Proof:} (i) and (ii) are equivalent by Corollary 2.7. (ii) and (iii)
are equivalent by Lemma 3.4. (iii) and (iv) are equivalent by the
following Theorem. \qed

(The equivalence (i) $\Leftrightarrow$ (iii) could have been
argued already from Lemma 2.3.)

{\bf 3.6.\ Theorem: \sl Let $A_L\otimes A_R\subset B$ be a CTPS with
  finite index, and $Z_{l,r}$ its coupling matrix, that is 
$$\rho=\bar\iota\comp\iota \simeq\bigoplus_{l,r} Z_{l,r}\;\rho^L_{l}
\otimes\rho^R_{r}$$
where $\iota: A_L\otimes A_R\to B$ denote the inclusion 
map and $\bar\iota$ its conjugate. If the coupling matrix satisfies
$$Z_{0,r}=\delta_{0,r}\quad {\rm and}\quad 
Z_{l,0}=\delta_{l,0}$$ 
(that is, the CTPS is normal and irreducible), then \\
(1) $Z$ is a permutation matrix. It induces a bijection $\,\hat\cdot\,$ 
with inverse $\,\check\cdot\,$ between the systems of sectors
$\{\rho^L_{l}\}$ and $\{\rho^R_{r}\}$ 
contributing to the decomposition of $\rho$ such that  
$$ Z_{l,r}=\delta_{\hat l,r}=\delta_{l,\check r}.$$
(2) Both systems of sectors $\{\rho^L_{l}\}$ and 
$\{\rho^R_{r}\}$ are closed under conjugation and under 
decomposition of products (fusion). They satisfy the same fusion rules
$$\rho^L_l\rho^L_k\simeq\bigoplus_m N_{lk}^m\;
\rho^L_m\qquad {\rm and}\qquad
\rho^R_r\rho^R_s\simeq\bigoplus_t \hat N_{rs}^t\; \rho^R_t$$
with
$ N_{lk}^m = \hat N_{\hat l\hat k}^{\hat m}$.
In particular, the bijection $\,\hat\cdot\,$ respects conjugation, and
the dimensions of the corresponding sectors coincide:
$$ d(\rho^R_{\hat l})=d(\rho^L_{l}).$$
(3) The homomorphisms $\iota^L_{l}:=\iota\comp(\rho^L_{l}\otimes\id_R): 
A_L\otimes A_R\to B$ are irreducible and mutually inequivalent. The
same holds for $\iota^R_{r}:=\iota\comp(\id_L\otimes\rho^R_{r})$, and
$\iota^R_{r}\simeq\iota^L_{\check r}$. Moreover, }
\begin{eqnarray*}\iota\comp(\rho^L_{l}\otimes\rho^R_{r}) \simeq\bigoplus_{k} 
N_{\check{\bar r}l}^{k}\; \iota^L_{k} \simeq\bigoplus_{s} 
\hat N_{\hat{\bar l}r}^{s}\; \iota^R_{s}. \end{eqnarray*} 

{\it Proof:} The proof adopts and extends methods taken from \cite{M}.

Let the index sets $\{l\}$ and $\{r\}$ label the irreducible sectors 
$\rho^L_{l}$ of $A_L$ and $\rho^R_{r}$ of $A_R$, respectively, obtained 
by closure under reduction of products of those sectors which contribute to 
$\rho$. If among these there are any ``new'' sectors not already
contributing to $\rho$, we extend the coupling matrix $Z$ by zero
columns and rows, but we are eventually going to show that there are 
no such new sectors. 

Only finitely many columns and rows of $Z$ are non-zero. Since 
$\rho=\bar\iota\comp\iota$ is self-conjugate, along with 
$\rho^L_{l}\otimes\rho^R_{r}$ also its 
conjugate must contribute with the same multiplicity, and hence 
$Z_{l,r}=Z_{\bar l,\bar r}$. In particular, both systems $\{\rho^L_{l}\}$
and $\{\rho^R_{r}\}$ are closed under conjugation.

Let the homomorphisms $\iota^L_{l}:A_L\otimes A_R\to B$ be as in (3). 
We compute 
\begin{eqnarray*} \langle \iota^L_{l},\iota^L_{l'}\rangle 
= \langle\iota\comp(\rho^L_{l}\otimes \id_R), 
\iota\comp(\rho^L_{l'}\otimes\id_R)\rangle  
= \langle \rho^L_{l}\otimes \id_R,
\bar\iota\comp\iota\comp(\rho^L_{l'}\otimes \id_R)\rangle = \qquad \\  
= \sum_{k,s} Z_{k,s} \langle \rho^L_{l}\otimes
\id_R,\rho^L_{k}\rho^L_{l'}\otimes \rho^R_{s}\rangle  
= \sum_{k,s} Z_{k,s} \langle \rho^L_{l},\rho^L_{k}\rho^L_{l'}\rangle 
\langle\id_R,\rho^R_{s}\rangle . \end{eqnarray*}
To this sum contributes only $s=0$ since $\langle \id_R,\rho^R_{s}\rangle 
=\delta_{s,0}$, and by the assumed properties of $Z$ also $k=0$ is the only
contribution. Hence
$$\langle \iota^L_{l},\iota^L_{l'}\rangle =
\langle \rho^L_{l},\rho^L_{l'}\rangle  = \delta_{l,l'}.$$
Thus the homomorphisms $\iota^L_{l}$ are irreducible and mutually 
inequivalent. The symmetric argument applies to $\iota^R_{r}$. Next we compute
\begin{eqnarray*} \langle \iota^L_{l},\iota^R_{\bar r}\rangle  = \langle 
\rho^L_{l}\otimes \id_R,\bar\iota\comp\iota\comp(\id_L\otimes \bar\rho^R_{r})
\rangle 
=\sum_{k,s} Z_{k,s} \langle \rho^L_{l},\rho^L_{k}\rangle \langle
\id_R,\rho^R_{s}\bar\rho^R_{r}\rangle = Z_{l,r}\; . \end{eqnarray*}
As we have seen that both sets of homomorphisms $\{\iota^L_{l}\}$ 
and $\{\iota^R_{r}\}$ consist of mutually inequivalent irreducibles, 
each $\iota^L_{l}$ can be
equivalent to at most one $\iota^R_{\bar r}$. Hence for fixed index $l$,
at most one entry $Z_{l,r}$ can be different from zero and must be
one. It follows also that no $\iota^L_{l}$ associated with a ``new'' sector 
$\rho^L_{l}$ can be equivalent to any of the $\iota^R_{r}$, old or
new, and vice versa.

For the ``old'' sectors, we write 
$$r=\hat l \quad\hbox{and}\quad l=\check r \qquad\hbox{iff}\quad
Z_{l,r}=1, \quad\hbox{that is, iff}\quad \iota^L_{l}\simeq\iota^R_{\bar r}.$$ 
That this assignment between old sectors is bijective follows from
transitivity of equivalence of sectors. Since we have already seen that $Z$ 
is conjugation invariant, this assignment respects conjugation, that is
$$\bar\rho^R_{\hat l}=\rho^R_{\bar{\hat l}}=\rho^R_{\hat{\bar l}}
\qquad {\rm etc.}$$

Next, we consider homomorphisms
$\iota_{l,r}:=\iota\comp(\rho^L_{l}\otimes\rho^R_{r}): A_L\otimes
A_R\to B$ and compute
\begin{eqnarray*} \iota_{l,r} =
\iota^R_{r}\comp(\rho^L_{l}\otimes \id_R)\simeq 
\iota^L_{\check{\bar r}}\comp(\rho^L_{l}\otimes \id_R)= 
\iota\comp(\bar\rho^L_{\check r}\rho^L_{l}\otimes \id_R) \simeq 
\bigoplus_{k} N_{\check{\bar r}l}^{k}\; \iota^L_{k}. 
\end{eqnarray*}
The symmetric argument produces also the decomposition
\begin{eqnarray*}\iota_{l,r}=\iota^L_{l}\comp(\id_L\otimes\rho^R_{r})
\simeq \iota\comp(\id_L\otimes\bar\rho^R_{\hat l}\rho^R_{r}) 
\simeq \bigoplus_{s} \hat N_{\hat{\bar l}r}^{s}\;\iota^R_{s}.
\end{eqnarray*}
In the first of these two decomposition formulae of the same object, no
``new'' label ${k}$ can appear, since we have seen that such a 
term $\iota^L_{k}$ is not equivalent to any term $\iota^R_{s}$ in 
the second decomposition formula, and vice versa. This shows that the
sets of sectors contributing to the coupling matrix are already closed
under reduction of products. Furthermore, comparison of the two
decomposition formulae shows equality of the multiplicities 
$N_{\check{\bar r}l}^{k}$ and $\hat N_{\hat{\bar l}r}^{s}\equiv
\hat N_{\bar r\hat l}^{\bar s}$ if $\bar s=\hat k$. Hence the
bijection $\,\hat\cdot\,$ between the sectors induces an isomorphism
of the fusion rules. 

Since finally the fusion rules of a finite system determine the
dimensions uniquely, also the equality of the dimensions follows. \qed

We have thus reproduced a result found previously in the 
classification program for modular invariant partition functions with 
heavy use of $\SL(2,\ZZ)$ machinery \cite{MS}, reducing every modular 
invariant to an ``automorphism of the fusion rules'' for suitably 
extended chiral observables. Our analysis is, however, much stronger 
since its assumptions are much weaker. Furthermore, it implies that the 
``suitably extended'' chiral observables are indeed the maximal chiral
observables defined in 2.1, and coincide with the relative 
commutants of the initially given chiral observables (Corollary 2.7(i)). 

Second, if possibly the maximal left and right chiral observables are not
isomorphic, then the result still implies an isomorphism of the
respective fusion rules. The corresponding statement is even more interesting 
in the case of coset models where typically $A\subset B$ is a theory
with well-known fusion rules, while the coset theory $C=A^c$ is in 
general a $W$-algebra whose superselection structure is a priori unknown. 
The theorem establishes that the fusion rules of this $W$-algebra are 
isomorphic to those of a local extension of the given theory $A$, 
namely the relative commutant $A^{cc}$ of $C$, which is in turn 
controllable in terms of the representations of $A$ itself. For coset
models based on current algebras, our result seems to be the algebraic
backbone of the modular reasoning as in \cite{SY}.

Finally, we emphasize that the sectors in Theorem 3.6 were never
referred to as being restrictions of DHR sectors. Neither was it
required that their fusion be abelian. The theorem is thus of a
quite more general nature than its specific application to conformal
quantum field theory as treated in this paper.

\section{Towards classification}

Modular invariant partition functions associated with affine Lie
algebras ($A_L \simeq A \simeq A_R$), as far as they have been 
classified, exhibit a classification scheme which refers to certain 
graphs and their exponents (eigenvalues of the square of the 
adjacency matrix) \cite{CIZ,G}. An essential statement is on the 
non-vanishing diagonal entries of the coupling matrix $Z$. 

A rather general formulation can be found in \cite[II]{BE}. 
It entails that $Z_{\lam,\lam}\neq 0$\footnote{In affine models the 
  DHR sectors of the initially given chiral observables are given in terms of
  weights $\lam$ of semisimple Lie algebras. Throughout this section, 
  we adopt the labels $\lam$ for DHR sectors in order to make the
  present generalizations more transparent.}  
if and only if the DHR sector $\lam$ of $A$ belongs to a set of
``exponents'' associated with the chiral extensions $A \subset A\ext$. 
The set of exponents is a subset of the sectors of $A$.  

By modular invariance, the sectors of $A$ label at the
same time also the irreducible representations of their own fusion
algebra, the modular matrix $S$ playing the role of a ``generalized
Fourier transformation'' between the fusion algebra itself and its dual. 
On the other hand, modular invariance of the partition function implies
that the coupling matrix coincides with its Fourier transform (up to a
conjugation). Hence, the above statement on the sector $\lam$ being an
exponent can as well be interpreted as a statement on the irreducible
representation $\lam$ of the fusion algebra and on non-vanishing 
entries of the Fourier transformed coupling matrix. In the following, 
we set out to 
formulate a generalization of this version of the statement to the
more general situation we discussed in this paper (without parity 
symmetry between left and right chiral algebras, and without
assumption of modular invariance).

Let $A_L\otimes A_R\subset A\max_L\otimes A\max_R\subset B$ denote some 
initially given chiral observables embedded into a two-dimensional local
theory $B$ (satisfying the assumptions of section 2) along with their 
maximal chiral extensions obtained by passing to the relative
commutants in $B$.

Let $W_L$ and $W_R$ denote the fusion algebras of all irreducible DHR
sectors $\lam_L$, $\lam_R$ of the initially given chiral observables (or
fusion subalgebras containing all sectors which contribute to the  
coupling matrix $Z$). Let $W\max_L$ and $W\max_R$ denote the fusion
algebras of the irreducible sectors $\tau_L$, $\tau_R$ of the extended
(= maximal) chiral observables which contribute to the coupling 
matrix (i.e., which are contained in the vacuum representation of $B$). 
According to Theorem 3.6 and Corollary 3.5, the fusion algebras 
$W\max_L$ and $W\max_R$ are isomorphic under the bijection
$\,\hat\cdot\,$. We use this bijection to identify $W\max_L$ with
$W\max_R$, so the coupling matrix with respect to $A\max$ becomes the 
unit matrix $\eins$. 
 
To be on safe grounds, we assume that $W_L$ and $W_R$ contain only finitely 
many sectors, and that these have finite dimensions. This implies the 
same for $W\max$, and ensures that all extensions have finite index.

Restriction and extension prescriptions between DHR sectors
of a theory $B$ and a subtheory $A$ were given in \cite{LR}, and
further analyzed in \cite{BE}. We are going to apply this theory to
the chiral extensions $A\max_L$ of $A_L$, and $A\max_R$ of $A_R$. 

The restriction is just the restriction of 
representations and coincides with the ``canonical'' prescription in 
terms of the inclusion homomorphism $\iota$ and its conjugate, given by
$\tau\mapsto\sig_\tau=\bar\iota\comp\tau\comp\iota$. 
It was named $\sig$-restriction in \cite{BE}. In the present situation,
$\sig$-restriction maps $W\max$ into $W$.\footnote{Here and in the
  sequel, we often suppress the subscripts $L$ and $R$ when both 
  chiralities are understood.}

In contrast, the extension prescription $\lam\mapsto\a_\lam$ \cite{LR}
differs from the canonical induction 
$\lam\mapsto\iota\comp\lam\comp\bar\iota$; it was named $\a$-induction
in \cite{BE} for distinction.
In particular, unlike canonical induction, $\a$-induction respects sector
composition, and the trivial sector of the subtheory extends to the
trivial sector of the extended theory. Furthermore, $\a$-extensions of
DHR sectors of the subtheory in general are not DHR but only
half-space localized (solitonic) sectors, due to a monodromy
obstruction \cite{LR}. Let $V_L$ and $V_R$ denote the, possibly
non-abelian, fusion algebras of all sectors (labelled $\beta$) generated by
reduction of products of $\a$-extended DHR sectors from $W_L$ and $W_R$.

In \cite{BE}, a reciprocity formula for $\a$-induction and
$\sig$-restriction was found:
$$\langle\a_\lam,\tau\rangle=\langle\lam,\sig_\tau\rangle$$
provided $\lam$ and $\tau$ are DHR sectors of the respective
theories. It entails that $\a_\lam$ and $\iota\comp\lam\comp\bar\iota$, 
while otherwise different, contain the same DHR subsectors.
It also entails that, in the present setting, the fusion algebras $V$ 
contain the abelian subalgebras $W\max$.

Let $B_L$ and $B_R$ denote the rectangular ``branching matrices'', describing 
chiral $\sig$-restriction, with non-negative integer entries 
$\langle\lam,\sig_\tau\rangle$ which connect the irreducible DHR
sectors $\tau\in W\max$ with $\lam\in W$. Then the (in general
rectangular, $\dim W_L \times \dim W_R$) coupling matrix
with respect to the initially given chiral observables is
$$Z=B_LB_R^t,$$
that is, $Z_{\lam_L,\lam_R}\neq 0$ if and only if the sectors $\lam_L$ and
$\lam_R$ arise by restriction from a pair of sectors of the maximal
chiral observables which are identified by the bijection $\,\hat\cdot\,$ of
Theorem 3.6. This is just the ``block form'' of the coupling matrix 
expected by restricting first $\pi^0_B$ to the maximal chiral
observables, and subsequently restricting the sectors so obtained to 
the initially given chiral observables.
 
Each fusion algebra has a ``regular representation'' defined by 
representing a sector by its matrix of fusion multiplicities with the other
sectors. $W$ and $W\max$ being abelian, all their irreducible representations
are one-dimensional and contribute with multiplicity one to the
regular representations. The values of the generators of the
fusion algebra in the irreducible representations provide ``character
tables'' which are non-degenerate square matrices. We denote the
one-dimensional representations of $W$ by $\phi\in\widehat W$, and
their character tables by $X$.

The character table defines a ``generalized Fourier transform'' between
any abelian fusion algebra and its representations. The Fourier transformed
coupling matrix is thus defined as
$$\widehat Z = (X_LB_L)(X_RB_R)^t.$$
Its matrix entries are the values of the restriction of the vacuum
sector of the 2D theory $B$, as a DHR sector of $A_L\otimes A_R$, 
in the irreducible representations $\phi_L\otimes\phi_R$ of the tensor
product $W_L\otimes W_R$ of the chiral DHR fusion algebras. A priori,
the entries need not to be integers.

Let $\bar\phi\in\widehat W$ denote the conjugate representation of $\phi$.
Since the adjoint in the fusion algebra is given by sector conjugation, 
we have $\phi(\bar\lam)=\overline{\phi(\lam)}=\bar\phi(\lam)$. This means 
$\widehat C X = XC=\overline{X}$ where $C$ and $\widehat C$ are
the conjugation matrices for the DHR sectors of the initially given chiral
observables $A$ and for the representations of their fusion algebras
$W$, respectively. Furthermore, restriction respects sector conjugation,
hence $BC\max=CB$ where $C\max$ is the conjugation matrix for the
sectors $\tau\in W\max$ of the maximal chiral observables $A\max$. 

Thus, since the branching matrices are real, we arrive at
$$\widehat Z\widehat C=(X_LB_L)(X_RB_R)^+\qquad\hbox{or equivalently}\qquad
\widehat Z=(X_LB_L)C(X_RB_R)^+.$$

It follows that a matrix entry of $\widehat Z\widehat C$ to be non-zero
requires that the corresponding complex row vectors of $X_LB_L$ and 
$X_RB_R$ are not orthogonal, and a fortiori non-zero. If both the
chiral branching and the chiral fusion algebras are isomorphic, e.g.,
if the theory $B$ is parity symmetric, then a diagonal matrix entry of 
$\widehat Z\widehat C$ vanishes if {\em and only if} the corresponding
row vector of $XB$ vanishes.

A modular (transformation) matrix $S$, if it exists, establishes a 
natural identification between the generators of a fusion algebra and
its representations, and $X=S$. Since $S^2=C=\widehat C$, 
modular $S$-invariance is the statement that the coupling matrix 
$Z=SZS^*=\widehat Z\widehat C$ equals its own Fourier transform up to
a conjugation. 

This remark implies that the Proposition 4.1 below 
reduces to the above-mentioned statement on ``exponents'' in \cite{BE}
in the case with modular invariance.

We have first to adapt definitions made in \cite[II]{BE} to our more
general setting. We introduce certain subsets of $\widehat W$ which reflect
the structure of the chiral extensions. 

For a given irreducible 
DHR sector $\tau\in W\max$, we define the {\bf $\sig$-supports}
$\Supp_L(\tau)$ and $\Supp_R(\tau)$ as the subsets of those irreducible 
representations of $W_L$ and $W_R$ which do not vanish on the
respective restrictions $\sig_\tau$ of $\tau$ to the initially given
left and right chiral observables, that is, those rows of $XB$ which have 
non-zero entry in column $\tau$. The notion ``support'' is motivated by
considering the abelian fusion algebra $W$ as an algebra of functions
on the set $\widehat W$ of its one-dimensional representations. 
Thus $\Supp(\tau)\subset\widehat W$ is indeed the support of the function 
$\sig_\tau\in W$. (The $\sig$-supports were called $\rm Eig(\tau)$
in \cite{BE}.)

As shown in \cite{BE}, $\a$-induction of sectors induces a
homomorphism of fusion algebras $W\to V$. Composing this homomorphism 
with the regular representation of $V$ yields another representation,
$\pi_\a$, of $W$. We define the {\bf $\a$-spectra} $\Spec_L$ and $\Spec_R$ 
as the subsets of those irreducible representations of $W_L$ and $W_R$
which are contained in the $\a$-induced representations $\pi^L_\a$ and
$\pi^R_\a$. (The $\a$-spectra were called $\rm Exp$  
in \cite{BE} and are the ``exponents'' mentioned above.)

Now, by virtue of $\a$-$\sig$-reciprocity \cite{BE}, we are going to
derive 

{\bf 4.1.\ Proposition: \rm (i) \sl A matrix entry of $\widehat Z\widehat C$
vanishes unless for some sector $\tau\in W\max$, both matrix indices
belong to the respective left and right $\sig$-supports $\Supp(\tau)$. 
It also vanishes unless both matrix indices belong to the left and 
right $\a$-spectra $\Spec$.  

\rm (ii) \sl If (fusion and branching of) the left and right chiral 
theories are 
isomorphic, then a diagonal matrix entry of $\widehat Z\widehat C$ is 
non-zero if and only if the corresponding representation of $W$
belongs to the union $\bigcup_\tau \Supp(\tau)$.}

In fact, there are many interesting cases when $\bigcup_\tau
\Supp(\tau)=\Spec$ (some of them being given below), so the last 
statement can be phrased in terms of the $\a$-spectrum $\Spec$. 

The Proposition is the desired generalization of the classification statement 
\cite{CIZ,G,BE} for modular invariant partition functions. 
(The second statement seems not to be
sensible with differing left and right chiral fusion and branching
matrices, since the product of two different row vectors can clearly 
vanish without these vectors being zero.)

The Proposition makes assertions about the coupling matrix for the
initially given  
chiral observables $A_L\otimes A_R$ embedded into the 2D theory $B$,
in terms of the chiral extensions $A\subset A\max$ to which
$\a$-induction and $\sig$-restriction pertain. Thus the 2D problem
is reduced to a chiral problem. An open issue remains, however, a
model-independent classification of possible $\a$-spectra, and hence 
of 2D chiral extensions. The available classifications for affine 
Lie and Virasoro algebras (``diagonal or automorphism, orbifold, 
exceptional'' \cite{CIZ,G,BE}) refer to the chiral extensions being in turn 
trivial, fixpoints under an abelian group, or conformal embeddings, 
and are expected to be too coarse in the general case.

{\it Proof of the Proposition:} (i) The first statement 
is obvious since by the
representation $\widehat Z\widehat C=(X_LB_L)(X_RB_R)^+$, every matrix
entry is the inner product of row vectors whose components are the 
values of the functions $\sig_\tau$, $\tau\in W\max$, evaluated on the
respective left and right
one-dimensional representations. The inner product vanishes whenever 
these representations do not belong to the respective $\sig$-supports.
The second statement is a consequence of the first in view of the
Lemma 4.2 below. \\
(ii) For isomorphic left and right chiral fusion and branching,
$X_LB_L=X_RB_R$,  
diagonal matrix entries of $\widehat Z\widehat C$ are 
norm squares of row vectors of $XB$ which vanish if {\it and only if} all
their entries vanish, hence if {\it and only if} the corresponding
representation of $W$ does not belong to any of the $\sig$-supports
$\Supp(\tau)$, $\tau\in W\max$. \qed

We have used

{\bf 4.2.\ Lemma: \sl $\bigcup_\tau \Supp(\tau)\subset \Spec$. }

{\it Proof:} The one-dimensional representations $\phi$ of an
abelian fusion algebra with generators $\lam$, considered as
vectors with entries $\phi(\lam)$, are pairwise orthogonal
\cite{K}. This property enables us to decide whether a representation 
$\phi$ is contained in the $\a$-induced representation $\pi_\a(\lam)$ with
matrix entries $\langle\a_\lam\b_1,\b_2\rangle$, by contracting
the matrix-valued vector $(\pi_\a(\lam))_\lam$ with the vector 
$(\overline{\phi(\lam)})_\lam$. Thus $\phi$ belongs to the $\a$-spectrum
$\Spec$ if and only if the resulting matrix 
$$(\overline\phi\cdot\pi_\a)_{\b_1\b_2}\equiv
\sum_\lam\overline{\phi(\lam)}\pi_\a(\lam)_{\b_1\b_2}
=\sum_\lam\overline{\phi(\lam)}\langle\a_\lam \b_1,\b_2\rangle$$
is non-zero. But for
$\b_1=\id_{A\max}$, and $\b_2=\tau$ an irreducible sector from 
$W\max\subset V$, the matrix entry of the $\a$-induced
representation equals $\langle\lam,\sig_\tau\rangle$ by
$\a$-$\sig$-reciprocity, and the contracted matrix entry equals
$\overline{\phi(\sig_\tau)}$. Hence, if $\phi$ belongs to any of the
$\sig$-supports $\Supp(\tau)$, then $\phi$ belongs to the
$\a$-spectrum $\Spec$. \qed 

We list here two ``extremal'', but by no means exhaustive, conditions
to ensure equality in Lemma 4.2, that is, $\bigcup_\tau \Supp(\tau) = \Spec$:

{\bf 4.3.\ Lemma: \sl If $\a$-induction is surjective 
(considered as a linear map from $W$ into $V$), 
then $\Supp(\id_{A\max})=\bigcup_\tau \Supp(\tau)=\Spec$. 

If $\sig$-restriction is surjective (considered as a linear map from $W\max$
into $W$), then $\bigcup_\tau \Supp(\tau) = \Spec = \widehat W$
exhaust all representations of $W$. }
 
The case of surjective induction was also paid special attention in \cite{BE}.
Indeed, there are many other cases when $\bigcup_\tau \Supp(\tau)=\Spec$, 
but we have no satisfactory characterization yet.

{\em Proof:} We want to compute the $\sig$-support $\Supp(\id_{A\max})$. 
For this purpose, we multiply $\phi(\sig_{\id_{A\max}})$ with 
$\phi(\mu)$, $\mu\in W$. Using in turn $\a$-$\sig$-reciprocity, 
the representation condition for $\phi$, Frobenius reciprocity, 
the homomorphism property of $\a$-induction, and associativity of
fusion, we arrive at
\begin{eqnarray*}\phi(\sig_{\id_{A\max}})\phi(\mu)=
\sum_\lam\phi(\lam)\phi(\mu)\langle\a_\lam,\id_{A\max}\rangle =
\sum_{\kappa\lam} N_{\kappa\bar\mu}^\lam\phi(\kappa)
\langle\a_\lam,\id_{A\max}\rangle =\\=
\sum_\kappa\phi(\kappa)\langle\a_\kappa\bar\a_\mu,\id_{A\max}\rangle=
\sum_{\kappa,\b}\phi(\kappa)\langle\bar\a_\mu,\b\rangle\langle 
\a_\kappa \b,\id_{A\max}\rangle.\end{eqnarray*}
Here the sum over $\b$ extends over all sectors of $V$. The last sum 
must vanish for every $\mu$, since the left hand side does, if 
$\phi(\sig_{\id_{A\max}})=0$,
i.e., if $\phi\not\in\Supp(\id_{A\max})$. Now, if $\a$-induction 
is surjective, then every sector $\b$ arises as a linear combination 
of sectors $\a_\mu$, and consequently
$$\sum_\kappa\phi(\kappa)\langle \a_\kappa \b,\id_{A\max}\rangle=
\sum_\lam\overline{\phi(\lam)}\langle \a_\lam,\b\rangle $$
must vanish for all $\b$. These are sufficiently many matrix
entries to ensure the vanishing of the full matrix (since $\langle\a 
\b_1,\b_2\rangle=\sum_\b\langle\a,\b\rangle\langle \b\b_1,\b_2\rangle$), 
and hence the absence of $\phi$ from the $\a$-spectrum $\Spec$. Hence
$\Spec\subset\Supp(\id_{A\max})$, implying the first claim.

On the other hand, if $\sig$-restriction is surjective, then
$\phi(\sig_\tau)=0$ for all $\tau\in W\max$ implies $\phi(\lam)=0$ 
for all $\lam\in W$, hence $\phi=0$. Thus the union of the 
$\sig$-supports exhausts all representations of $W$, implying the
second claim. \qed

We have thus established some first constraints on the coupling matrix 
in terms of representations of fusion algebras.

Further constraints are expected to derive from locality which was
only partially exploited in the form of $\a$-$\sig$-reciprocity in
Proposition 4.1, and in the commutativity of left and right chiral
observables in Theorem 3.6. Notably the condition for locality of the 2D theory
in terms of the local subfactor data and the statistics which was given 
in \cite{LR} remains to be transcribed into a condition on 
the coupling matrix.

As mentioned in the introduction, chiral locality produces matrices
$S\stat$ and $T\stat$ which represent $\SL(2,\ZZ)$ 
\cite{Pal,FG}, except for a possible degeneracy of the braiding.
A first implication of the locality condition for the 2D theory is that 
$T\stat_LZ=ZT\stat_R$, in accordance with local 2D conformal fields
having integer spin $h_L-h_R$. The companion relation
$S\stat_LZ=ZS\stat_R$, that is, modular invariance of the coupling
matrix with respect to the representation of $\SL(2,\ZZ)$ given by the
statistics, cannot be established for general 2D nets $B$,
however. The surprise is that, as shown here, one can go much of the
way towards classification without knowing these formulae, and that
one can do so whether the involved sectors have a degenerate braiding or
not. (M\"uger's proof \cite{MM} that the degeneracy can always be
removed by an algebraic extension of the chiral observables does not
help here, since this extension is in general not possible within the
given 2D observables.) 

\section{Conclusions}

We have shown that in a 2D conformally invariant quantum field theory
with sufficiently many chiral observables to generate the chiral
M\"obius groups, there are maximal algebras of chiral observables
which are, locally, the relative commutants of each other, as well as
of any a priori given chiral observables sharing the generating
property (cf.\ Sect.\ 2).

The representation theory of the chiral observables is governed by a 
``canonical tensor product subfactor'' (CTPS) $A_L\otimes A_R\subset B$
given by the respective chiral and 2D local algebras. We have
therefore investigated the general structure of CTPS's and have found 
a characterization of the two tensor factors being each other`s relative 
commutants (``normality'') in terms of a coupling matrix.
The coupling matrix in this case provides an isomorphism between the
respective fusion rules for the involved sectors of the two tensor factors.

This abstract result, applied to the quantum field theoretical
situation at hand, generalizes a statement on certain ``extended'' 
chiral observables in the classification program for 2D modular 
invariant partition functions, and shows that the latter coincide with 
the maximal chiral observables.  

Exploiting general properties of $\a$-induction and $\sig$-restriction
between the superselection sectors of the maximal and the a priori 
given non-maximal chiral observables, constraints on the coupling 
matrix (with respect to the non-maximal chiral observables) are
derived which are the direct counterparts of similar constraints in 
the modular classification program.

Yet, modular invariance has not been assumed throughout the analysis. 
This supports our conviction that modular invariants are just one aspect
of a deeper and more general mathematical structure (presumably related 
to ``asymptotic subfactors'' and ``quantum doubles''). A classification 
in terms of graphs still remains to be established in the general 
situation. Possibly, additional constraints originating from locality 
will play a role here.

\vskip 10mm 

{\bf Acknowledgements:} I am indebted to J. B\"ockenhauer for many
helpful and critical comments on an earlier version of this paper, as
well as to A. Recknagel for discussions on modular invariance. I also
thank D. Buchholz, B. Schroer and H.-W. Wiesbrock for useful 
suggestions concerning Sect.\ 2, and R. Longo for pointing out a
difficulty in Sect.\ 3.


\small

\begin{thebibliography}{99} \itemsep=1pt
\bibitem{BE} J. B\"ockenhauer, D.E. Evans: {\em Modular invariants,
    graphs and $\a$-induction for nets of subfactors}, 
    I: Commun.\ Math.\ Phys.\ {\bf 197} (1998) 361-386, II: {\it ibid.}\ 
    {\bf 200} (1999) 57-103, and III: preprint hep-th/9812110 (1998).
\bibitem{BGL} R. Brunetti, D. Guido, R. Longo: {\em Modular structure
    and duality in conformal quantum field theory}, Commun.\ Math.\
    Phys.\ {\bf 156} (1993) 201-219.
\bibitem{CIZ} A. Cappelli, C. Itzykson, J.-B. Zuber: {\em The A-D-E
    classification of minimal and $A_1^{(1)}$ conformal invariant
    theories}, Commun.\ Math.\ Phys.\ {\bf 113} (1987) 1-26.
\bibitem{C} J.L. Cardy: {\em Operator content of two-dimensional
    conformally invariant theories}, Nucl.\ Phys.\ {\bf B270} (1986) 186-204.
\bibitem{FRS} K. Fredenhagen, K.-H. Rehren, B. Schroer: {\em
    Superselection sectors with permutation group statistics and
    exchange algebras}, I: Commun.\ Math.\ Phys.\ {\bf 125} (1989)
  201-226, and II: Rev.\ Math.\ Phys.\ {\bf Special Issue} (1992) 113-157. 
\bibitem{FG} J. Fr\"ohlich, F. Gabbiani: {\em Braid statistics in
    local quantum theory}, Rev.\ Math.\ Phys.\ {\bf 2} (1991) 251-353.
\bibitem{FG2} J. Fr\"ohlich, F. Gabbiani: {\em Operator algebras and
    conformal field theory}, Commun.\ Math.\ Phys.\ {\bf 155} (1993) 569-640.
\bibitem{G} T. Gannon: {\em The classification of affine SU(3) modular
    invariants}, Commun.\ Math.\ Phys.\ {\bf 161} (1994) 233-264.
\bibitem{GL} D. Guido, R. Longo: {\em The conformal spin and
    statistics theorem}, Commun.\ Math.\ Phys.\ {\bf 181} (1996) 11-36.
\bibitem{GLRV} D. Guido, R. Longo, J.E. Roberts, R. Verch: {\em A
    general framework for charged sectors in quantum field theory on
    curved spacetimes}, preprint in preparation.
\bibitem{LQP} R. Haag: {\em Local Quantum Physics}, Springer (1992). 
\bibitem{VK} V.G. Kac: {\em The idea of locality}, in: ``Group 21'',
     Proceedings Goslar 1996, H.-D. Doebner et al eds.\ (World Scientific
     1997), pp.\ 16-32.
\bibitem{KP} V.G. Kac, D.H. Peterson: {\em Infinite-dimensional Lie
    algebras, theta functions and modular forms}, Adv.\ Math.\ {\bf
    53} (1984) 125-264.
\bibitem{K} T. Kawai: {\em On the structure of fusion algebras},
  Phys.\ Lett.\ {\bf B 217} (1989) 247-251.
\bibitem{KLM} Y. Kawahigashi, R. Longo, M. M\"uger: {\em 
    Multi-interval subfactors and modularity of representations
    in conformal field theory}, preprint math.OA/9903104 (1999).
\bibitem{L} R. Longo: {\em Index of subfactors and statistics of
    quantum fields}, II: Commun.\ Math.\ Phys.\ {\bf 130} (1990) 285-309. 
\bibitem{LR} R. Longo, K.-H. Rehren: {\em Nets of subfactors}, Rev.\
    Math.\ Phys.\ {\bf 7} (1995) 567-597.
\bibitem{Dim} R. Longo, J.E. Roberts: {\em A theory of dimension},
  K-Theory {\bf 11} (1997) 103-159.
\bibitem{LM} M. L\"uscher, G. Mack: {\em Global conformal invariance
    in quantum field theory}, Commun.\ Math.\ Phys.\ {\bf 41} (1975) 203-234.
\bibitem{M} T. Masuda: {\em An analogue of Longo's canonical
    endomorphism for bimodule theory and its application to asymptotic
    inclusions}, Int.\ J. Math.\ {\bf 8} (1997) 249-265.
\bibitem{MM} M. M\"uger: {\em On charged fields with group symmetry
    and degeneracies of Verlinde's matrix $S$}, hep-th/9705018, 
    to appear in Ann.\ Inst.\ H. Poincar\'e.
\bibitem{MS} G. Moore, N. Seiberg: {\em Naturality in conformal
    field theory}, Nucl.\ Phys.\ {\bf B313} (1989) 16-40.
\bibitem{N} W. Nahm: {\em A proof of modular invariance}, Int.\ J.\
  Mod.\ Phys.\ {\bf A6} (1991) 2837-2845.
\bibitem{O} A. Ocneanu: {\em Quantum symmetry, differential geometry,
    and classification of subfactors}, Univ.\ Tokyo Seminary Notes
  {\bf 45} (1991) (notes taken by Y. Kawahigashi). 
\bibitem{Spt} K.-H. Rehren: {\em Space-time fields and exchange fields},
    Commun.\ Math.\ Phys.\ {\bf 132} (1990) 461-483.
\bibitem{SF} K.-H Rehren: in preparation.
\bibitem{Pal} K.-H. Rehren: {\em Braid group statistics and their
    superselection rules}, in: ``The algebraic theory of superselection
  sectors'', Proceedings Palermo 1989, D. Kastler ed.\ (World Scientific
  1990) pp.\ 333-355.
\bibitem{EA} K.-H. Rehren, B. Schroer: {\em Einstein causality and
    Artin braids}, Nucl.\ Phys.\ {\bf B312} (1989) 715-750.
\bibitem{SY} A.N. Schellekens, S. Yankielowicz: {\em Simple currents,
    modular invariants, and fixed points}, Int.\ J.\ Mod.\ Phys.\ 
    {\bf A 5} (1990) 2903-2952.
\bibitem{SW} B. Schroer, H.-W. Wiesbrock: {\em Looking behind the
    thermal horizon: hidden symmetries in chiral models}, preprint
  hep-th/9901031 (1999).)
\bibitem{TT} M. Takesaki: {\em Conditional expectations in von Neumann
    algebras}, J. Funct.\ Anal.\ {\bf 9} (1972) 306-321.
\bibitem{V} E. Verlinde: {\em Fusion rules and modular transformations
    in 2D conformal field theories}, Nucl.\ Phys.\ {\bf B300} (1988) 360-376.
\bibitem{HWW} H.-W. Wiesbrock: {\em Conformal quantum field theory and
    half-sided modular inclusions of von  Neumann algebras}, Commun.\ 
    Math.\ Phys.\ {\bf 158} (1993) 537-544.
\bibitem{X1} F. Xu: {\em Jones-Wassermann subfactors for disconnected
    intervals}, preprint q-alg/9704003 (1997).
\bibitem{X2} F. Xu: {\em Algebraic coset conformal field theories},  
    preprint math.OA/9810035 (1998).

\end{thebibliography}
\end{document}